\documentclass[twocolumn]{aastex62}

\graphicspath{{./}{figures/}}

\received{}
\revised{}
\accepted{}
\submitjournal{ApJ}

\shorttitle{Ser-emb 1}
\shortauthors{Mart\'in-Dom\'enech et al.}

\begin{document}

\title{A new, rotating hot corino in Serpens}

\correspondingauthor{Rafael Mart\'in-Dom\'enech}
\email{rafael.martin\_domenech@cfa.harvard.edu}

\author{Rafael Mart\'in-Dom\'enech}
\affil{Center for Astrophysics $|$ Harvard \& Smithsonian\\
60 Garden St., Cambridge, MA 02138, USA}

\author{Jennifer B. Bergner}
\affiliation{Harvard University Department of Chemistry and Chemical Biology \\
Cambridge, MA 02138, USA}

\author{Karin I. \"Oberg}
\affil{Center for Astrophysics $|$ Harvard \& Smithsonian\\
60 Garden St., Cambridge, MA 02138, USA}

\author{Jes K. J{\o}rgensen}
\affiliation{Niels Bohr Institute \& Centre for Star and Planet Formation, University of Copenhagen \\
{\O}ster Voldgade 5-7, DK-1350 Copenhagen K.}

\begin{abstract}
We 
have observed 29 transitions 
corresponding to 12 distinct species and 7 additional isotopologs toward the deeply embedded Class 0 young stellar object Ser-emb 1 
with the Atacama Large Millimeter Array at $\sim$ 1 mm. 
The detected species include CH$_3$OH and two complex organic molecules, CH$_3$OCH$_3$ and CH$_3$OCHO. 
The emission of CH$_3$OH and the two COMs is compact, and the CH$_3$OH rotational temperature is 261 $\pm$ 46 K,  
implying that Ser-emb 1 hosts a hot corino. 
The derived CH$_3$OH, CH$_3$OCH$_3$ and CH$_3$OCHO column densities are at least (1.2 $\pm$ 0.4) $\times$ 10$^{17}$ cm$^{-2}$, (9.2 $\pm$ 3.8) $\times$ 10$^{16}$ cm$^{-2}$, and (9.1 $\pm$ 3.6) $\times$ 10$^{16}$ cm$^{-2}$, respectively, 
comparable to the values found for other Class 0 hot corinos. 
%
In addition, 
we observe evidence of rotation at compact scales: 
two of the more strongly detected lines, corresponding to C$^{18}$O and H$_2$CO
present spatially resolved red- and blue-shifted compact emission orthogonal to 
the direction of a jet and outflow traced by CO, SiO, and several other molecules. 
The spatial coincidence of the hot corino emission and a possible disk in a compact region around the central protostar suggests that  
these structures may be physically and/or chemically related.

\end{abstract}


\section{Introduction} \label{sec:intro}
In the early stages of star formation \citep[i.e., Class 0/I sources, see, e.g.,][]{andre93,robitaille06}  the central protostar is embedded in a dense envelope as the parental dense core gravitationally collapses. 
This cold dense envelope warms up close to the central protostar and, where the temperature is high enough, the ice mantles on the surfaces of dust grains sublimate. This leads to the formation of a chemically-rich hot core (around high-mass protostars) or hot corino (around low-mass protostars\footnote{Hot corinos are thought to be similar to the pre-solar nebula that led to the formation of our own Solar System.}),  although  
not all observed warm compact cores present a hot core/corino chemistry \citep[see][and references therein]{herbst09}.  
Hot cores and hot corinos are characterized by the presence of saturated, complex (6+ atoms)  organic molecules (COMs) with abundances up to 10$^{-7}$ with respect to H$_2$, and rotational temperatures above 100 K (i.e., the water ice sublimation temperature). 
The typical size of hot corinos is about 100 au or less, with temperatures of around 100 K, while hot cores are warmer  (T $\sim$300 K) and much larger \citep[up to 10000 au,][]{herbst09}. 
Hot corinos are harder to detect than their more massive analogs, due to their lower envelope masses and luminosities  leading to weaker lines, and  
only nine hot corinos have been detected to date: IRAS 16293-2422 \citep{cazaux03}, 
IRAS 4A \citep{bottinelli04}, IRAS 2 \citep{jorgensen05}, IRAS 4B \citep{bottinelli07}, HH 212 \citep{codella16}, B335 \citep{imai16}, L483 \citep{oya17}, B1b \citep{lefloch18}, and SVS13-A \citep[the only Class I hot corino,][]{bianchi19}. 

COM observations are not limited to hot cores and hot corinos. 
\citet{marcelino07} reported the detection of 
methanol and acetaldehyde in the cold prestellar core TMC-1, with abundances up to only a few $\times$ 10$^{-9}$; and more recently, COMs have been also observed in the prestellar core L1544 \citep{vastel14,izaskun16}. 
In star-forming regions, COMs have been also detected in outflows
\citep{arce08,codella17,lefloch18}; and in the lukewarm and cold envelope around the B1b protostar \citep{oberg10}. 
These observations are consistent with a three-phases COM-formation scenario \citep[see, e.g.,][]{herbst09}. 
Roughly, during the cold phase of star formation (T $\sim$ 10 K, prior and during the isothermal collapse of the cold prestellar core), 
organics such as H$_2$CO and CH$_3$OH are mainly formed in the ice mantles on top of dust grains. 
The energetic processing of the ice mantles produce radicals such as HCO$^{.}$ and CH$_3$O$^{.}$ that are able to diffuse at lukewarm temperatures (T $>$ 20 K, once the protostar warms up the surrounding envelope) and react to form first-generation COMs like CH$_3$OCH$_3$ or CH$_3$OCHO \citep{garrod06,garrod08}, while   
further gas-phase reactions lead to the formation of the second generation of COMs during the hot core/corino phase. 

During the Class 0/I stages of star formation where hot corinos are detected, protostellar disks begin to form around the low-mass protostars due to conservation of angular momentum of the infalling-rotating envelope. 
Observations seem to show an evolutionary trend in protostellar disks \citep{yen13} from slow 
pseudodisks (a flattened envelope),
to Keplerian disks (disks rotating fast enough to become centrifugally supported). 
These disks further evolve to the protoplanetary disks observed around Class II/III T Tauri and Herbig Ae/Be protostars. 
%
Protostellar disks, and especially their chemistry, are harder to observe than the more evolved protoplanetary disks, 
mainly because of the difficulties in separating the emission from the disk and from the dense envelope in which it is embedded. 
Only a handful of protostellar disks have been confirmed toward Class 0 sources  \citep[namely, IRAS 4A2, L1527, VLA 1623A, L1448-mm, HH 212, IRS 7B, and Lupus 3MMS, see][] {choi10,tobin12,murillo13,yen13,codella14,lindberg14,yen17}, 
with sizes around 50$-$150 au, similar to the typical size of a hot corino. 

Previous works have tried to connect the presence of a hot corino with disk formation in Class 0/I sources. 
\citet{jorgensen05} had suggested a protostellar disk as a more dynamically stable location for the complex molecules in the IRAS 2A hot corino than the infalling envelope. 
Indeed, two of the nine detected sources with hot corinos (IRAS 4A2 and HH 212) show strong evidences of a protostellar disk, while the presence of such structure in a third hot corino source (IRAS 16293A) is also plausible. 
In IRAS 4A2, the NH$_3$ emission distribution from red- to blue-shifted velocities in the direction perpendicular to the bipolar jet was interpreted as a protostellar disk upon the fitting of the position-velocity (PV) diagram to a model of a disk with Keplerian rotation \citep{choi10}. %
A similar signature consistent with Keplerian rotation was found for the C$^{17}$O line emission in HH 212  \citep{codella14}. 
More recently, \citet{lee17,codella18,lee19} have observed emission from a total of 9 COMs above and below the edge-on dusty disk traced by the continuum emission in HH 212, indicating that the hot corino in this source is actually a warm atmosphere of the disk.
In other sources, though, the connection between hot corino emission and the protostellar disk is more tenuous.
\citet{oya16} and \citet{oya17} have studied the kinematic structure of two other sources presenting hot corino chemistry: IRAS 16293-2422 and L483, respectively. 
They use a three-dimensional ballistic model \citep{oya14} considering two physical components (a flattened infalling-rotating envelope and a Keplerian disk, separated by a centrifugal barrier) to simulate the emission of different molecular lines and reproduce their PV diagrams. 
In IRAS 16293A, they find that CH$_3$OH and CH$_3$OCHO are present in a ring-like structure around the centrifugal barrier with an inner radius of $\sim$55 au, interpreted as the evaporation from dust grains due to the accretion shocks expected in that region \citep{oya16}. 
Modelling of L483 in \citet{oya17} also indicates that the most likely location of the two detected COMs (NH$_2$CHO and CH$_3$OCHO) is the Keplerian disk formed within the centrifugal barrier. 
However, \citet{jacobsen18} ruled out the presence of a Keplerian disk in L483 down to at least 15 au, using higher spatial resolution observations, and concluded that the hot corino chemistry in this source is taking place at larger scales in the infalling-rotating envelope.



In this work, we explore the connection between hot corino emission and rotation in the Class 0 source Ser-emb 1.
The Serpens molecular cloud is a star-forming cloud harboring 34 embedded protostars \citep[9 Class 0 sources and 25 Class I sources,][]{enoch09}, located at a distance of $d$ = 436 $\pm$ 9 pc \citep{gisela18}. 
\citet{enoch11} reported evidences for a compact disk component around seven of them (Ser-emb 1, Ser-emb 4, Ser-emb 6, Ser-emb 7, Ser-emb 8, Ser-emb 11, Ser-emb 15, and Ser-emb 17) based on the continuum emission detected at long \textit{uv} baselines (and therefore coming from dust in compact regions around the central object). Among these seven sources, Ser-emb 1 is located in the Serpens Cluster B, and it is the least evolved protostar in the cloud according to its bolometric temperature (39 K). Its remaining envelope mass is 3.1 $M_{\odot}$\footnote{Class 0 and Class I protostars have accreted less than half its final mass, and therefore $M_{\star}$ $<$ $M_{env}$.}, and the estimated disk mass is 0.28 $M_{\odot}$ \citep{enoch11}. However, spatially and spectrally resolved line observations are needed to confirm the presence of a rotationally supported disk.

We present ALMA Band 6 observations of Ser-emb 1 aimed at 
the study of the structure of the source, and the evaluation of its chemical richness.  
The paper is organized as follows: 
ALMA observations, calibration, and imaging strategies are described in Sect. \ref{sec:observations}. 
An overview of the different structural components observed in Ser-emb 1 is presented in Sect. \ref{overview}. The molecular detections are presented in Sect. \ref{hotcorino}; and the evidences for a hot corino and a possible protostellar disk are shown in Sections \ref{spatial} and \ref{disk}, respectively, and discussed in Sect. \ref{discussion}. Section  \ref{conclusions} presents the final conclusions.

\section{Observations} \label{sec:observations}

The Ser-emb 1 observations analyzed in this paper were obtained within a larger ALMA project (\#2015.1.00964.S) 
on chemistry in disks at different evolutionary stages. 
The observations were centered at $RA_{\rm J2000}$ = 18h 29min 09.09s, $DEC_{\rm J2000}$ = +00$^{\circ}$ 31$^{\prime}$ 30.90$^{\prime\prime}$, and completed during Cycle 3 using two different Band 6 frequency settings. 
The first frequency setting consisted on eight spectral windows placed between 217 GHz and 233 GHz. 
Seven spectral windows have bandwidths of 58.6 or 117.2 MHz, and a spectral resolution of $\sim$0.16 km s$^{-1}$. 
The eighth spectral window has a bandwidth of 1.875 GHz, and a lower spectral resolution of $\sim$0.63 km s$^{-1}$.  
This frequency setting was observed on June 05 and 14, 2016, using 42 antennas (longest baseline of 772.8 m) for a total on source time of 19.2 min.   
The second frequency setting included seven spectral windows placed between 243 GHz and 262 GHz. 
Six spectral windows have a bandwidth of 117.2 MHz, and a spectral resolution of $\sim$0.14 km s$^{-1}$. 
The seventh spectral window has a bandwidth of 1.875 GHz, and a spectral resolution of $\sim$0.60 km s$^{-1}$. The observations of this frequency setting were carried out on May 15 and 16, 2016, using 41 antennas (longest baseline of 640.0 m) for a total on source time of 22.5 min.  
The observed spectral windows are listed in the Appendix. 

\subsection{Calibration}
The observed visibilities were initially calibrated by ALMA staff with the Common Astronomy Software Applications (CASA) versions 4.5.3 and 4.7.0, using the sources J1751+0939, J1830+0619 as bandpass and phase calibrators, respectively, and Titan as the absolute flux calibrator. 
The continuum data of each spectral window (channel-averaged visibilities after flagging line emission channels) was further self-calibrated in 1-2 rounds, and the solutions were applied to the native resolution visibilities. 
The continuum was substracted from the self-calibrated visibilities in the $uv$ plane using line-free channels. 

\subsection{Generation of the source spectrum}\label{obs_spec}
Imaging of the self-calibrated, continuum-substracted data cubes at their native spectral resolution 
was performed with the CASA version 5.3.0 using the task TCLEAN with Brigss weighting of the baselines. 
During the imaging process, an auto-mask was applied independently to every channel of the self-calibrated data cubes, using the parameters \textit{sidelobethreshold} = 2.5 $\times$ rms, \textit{noisethreshold} = 3.0 $\times$ rms, \textit{minbeamfrac} = 0.3, \textit{cutthreshold} = 0.4, lownoisethreshold = 1.5 $\times$ rms, and \textit{growiterations} = 75. 
For confirmation purposes, we imaged the C$^{18}$O 2 $-$ 1 line manually masking every channel, and obtained the same result.

The robustness parameter was set to 0.5 for the Briggs weighting of the baselines during this first round of imaging. 
The rms per channel (averaged over ten line-free channels) is 5$-$9 mJy beam$^{-1}$ in the high-spectral resolution windows, and $\sim$2 mJy beam$^{-1}$ in the low-spectral resolution windows. 
The average synthesized beam was 0.57$\arcsec$ $\times$ 0.47$\arcsec$. 
The particular values for each spectral window can be found in Sect. \ref{app-obs}, where the exact frequency range covered and spectral resolution for each spectral window is also listed. 
The spectrum of the source at the central pixel, which corresponds to the position of the continuum peak, was then extracted from the FITS file generated by CASA for every cleaned data cube, and subsequently used to detect and identify molecular transitions (see Sect. \ref{hotcorino} and Fig. \ref{spec}).

\subsection{Imaging of the individual lines}\label{obs_im}
The identified transitions 
in the extracted spectrum (see Fig. \ref{spec} and Table \ref{lines1}) were individually re-imaged following the same process explained above, but using a robustness parameter of 0.0 to improve the angular resolution. The averaged rms of the image cubes per channel increased to 6 $-$ 12 mJy beam$^{-1}$ ($\sim$3 mJy beam$^{-1}$ for the low-resolution spectral window), while the average size of the synthesized beam decreased to $\sim$ 0.50$\arcsec$ $\times$ 0.42$\arcsec$ (values for each spectral window can be found in Table \ref{spw}), which translates to a beam radius of $\sim$100 au at a distance of 436 pc.  
Two of the targeted lines, N$_2$D$^+$ 3 $-$ 2 and C$_2$H 3 $-$ 2, 
were not observed in the extracted spectrum, but extended emission is present.  
These lines were imaged with a robustness parameter of 0.5 to enhance the signal-to-noise ratio. 
All images were corrected for the primary beam. 
%

\section{Results} \label{sec:results}

\begin{figure*}
\centering
\includegraphics[width=14cm]{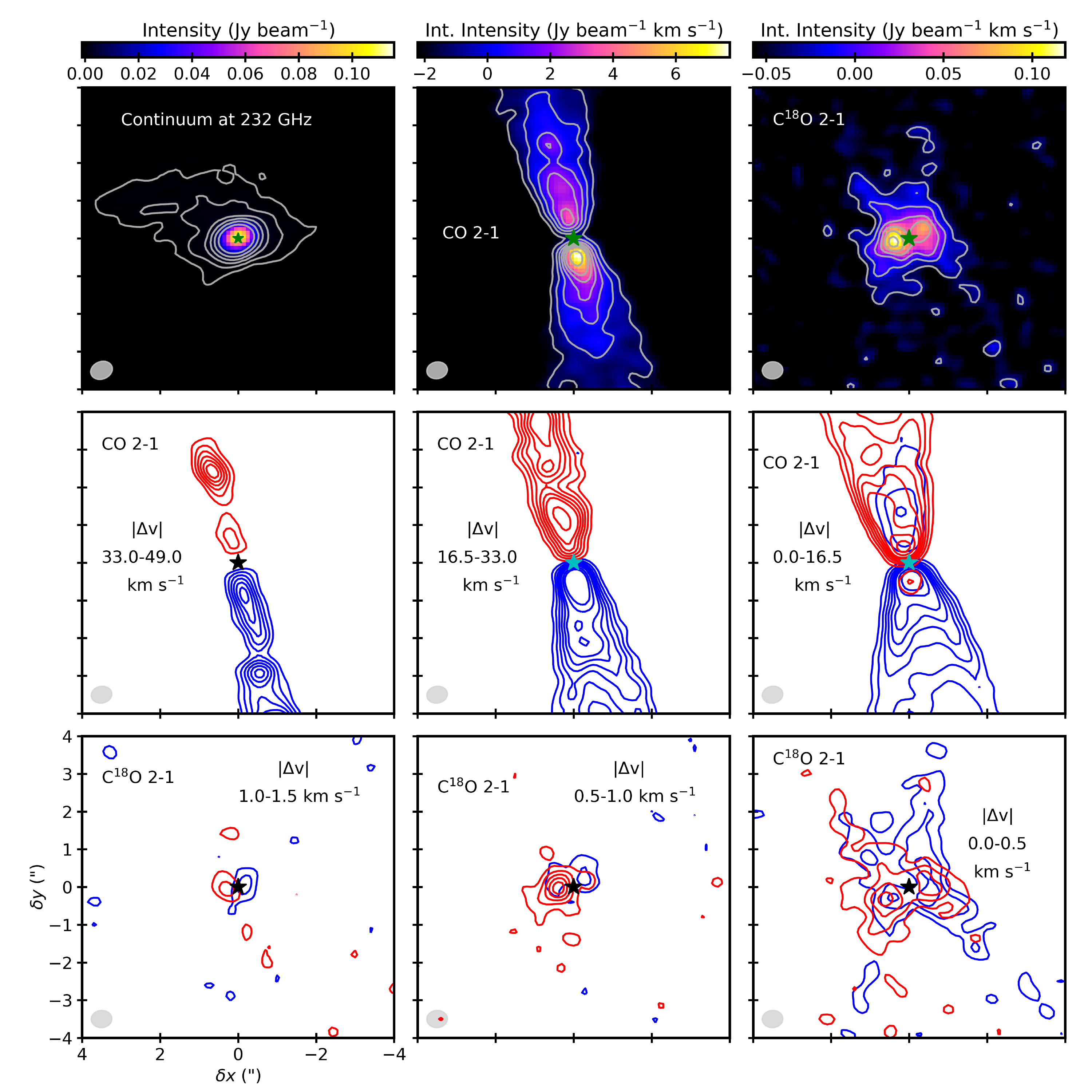}
\caption{The top left panel shows the continuum map centered at 232 GHz along with the 3, 6, 12, 18, 24, 48, and 96$\sigma$ contours ($\sigma$ = 0.3 mJy beam$^{-1}$). The position of the continuum peak is indicated with a star in every panel. The middle and right top panels show the moment 0 maps along with the 3, 6, 9, 12, 15, 18, 24, and 30$\sigma$ contours for the CO and C$^{18}$O 2 $-$ 1 transitions, respectively (maps are clipped at 1$\sigma$). The moment 0 rms is indicated in Table \ref{lines1}. The blue- and red-shifted moment 0 maps, split into three different velocity components are shown in the middle and bottom rows (only contours). The velocity range of every velocity component is indicated in every panel, and the size of the synthesized beam ($\sim$ 0.50$\arcsec$ $\times$ 0.42$\arcsec$) is also shown in the lower left of every panel (see also Sect. \ref{app-obs})}
\label{overview2}
\end{figure*}

\subsection{Overview of the source\label{overview}}

Figure \ref{overview2} presents an overview of the Ser-emb 1 source structure, as traced by millimeter dust continuum emission, and CO 2 $-$ 1 and C$^{18}$O 2 $-$ 1 line emission. 
The continuum map shown in the upper left panel was generated by combining all line-free channels in the wide-band spectral window centered at $\sim$232 GHz.  
The continuum peak is assumed to correspond to the position of the central protostar, and is marked in the remaining panels. 

The integrated moment 0 map over the full line width of the CO 2 $-$ 1 emission is shown in the upper middle panel of Fig. \ref{overview2}, and traces a bipolar outflow in the north-east to south-west direction (NE$-$SW, position angle of PA = 12.2$^{\circ}$) driven by the central protostar. 
The integrated blue- and red-shifted emission with respect to the systemic velocity of the source ($V_{LSR}$ $\sim$ 8.5 km s$^{-1}$) is represented in the middle row panels of Fig. \ref{overview2}, 
decomposed into three velocity components.  
The high-velocity emission ($|\Delta$v$|$ $>$ 33 km s$^{-1}$ with respect to $V_{LSR}$ of the source, middle left panel) traces a collimated jet, while the low-velocity emission ($|\Delta$v$|$ $<$ 16 km s$^{-1}$, middle right panel) shows a wide-angle, v-shaped slower outflow. 

The top right panel of Fig. \ref{overview2} shows the integrated moment 0 map of the C$^{18}$O 2 $-$ 1 emission, 
which is limited to the core of the source and the slower outflow region traced by the low-velocity CO 2 $-$ 1 emission. 
The three bottom panels present the integrated blue- and red-shifted C$^{18}$O emission, also split into high- ($|\Delta$v$|$ $>$ 1.0 km s$^{-1}$), mid- (0.5 $<$ $|\Delta$v$|$ $<$ 1.0 km s$^{-1}$), and low-velocities ($|\Delta$v$|$ $<$ 0.5 km s$^{-1}$, from left to right, respectively) components. We note that these three velocity components are all included in the low-velocity map of the much broader CO emission. 
The low-velocity component (bottom right panel) presents emission detected above the 3$\sigma$ level up to $\sim$2 $\arcsec$ from the central protostar ($\sim$800 au). 
The blueshifted emission is brighter on the west side, while the redshifted emission is predominantly seen in the east side of the source. A similar distribution for the HCO$^{+}$ 4 $-$ 3 low velocity ($|\Delta$v$|$ $<$ 1.5 km s$^{-1}$) emission in HH 212 was interpreted as a rotation contribution to the infall motion of the flattened envelope in \citet{lee17}. 
The high-velocity component (bottom left panel) is detected in a compact region within $\sim$0.5$\arcsec$ ($\sim$200 au) of the central protostar, and is thus marginally resolved with a beam size of $\sim$ 0.53$\arcsec$ $\times$ 0.46$\arcsec$ (Table \ref{spw}). 
Therefore, the actual radius of this compact component could be $<$200 au. 
The emission 
shows a gradient from red- to blue-shifted velocities in the direction perpendicular to the CO jet (PA = -77.7$^{\circ}$), indicating that rotation may dominate the motion of this component \citep[see, e.g.,][]{lee17}. 
The rotation could not be checked for the CO 2 $-$ 1 transition because of self-absorption at low velocities close to the continuum peak.

\subsection{Molecular detections\label{hotcorino}}

\begin{figure*}
\centering
\plotone{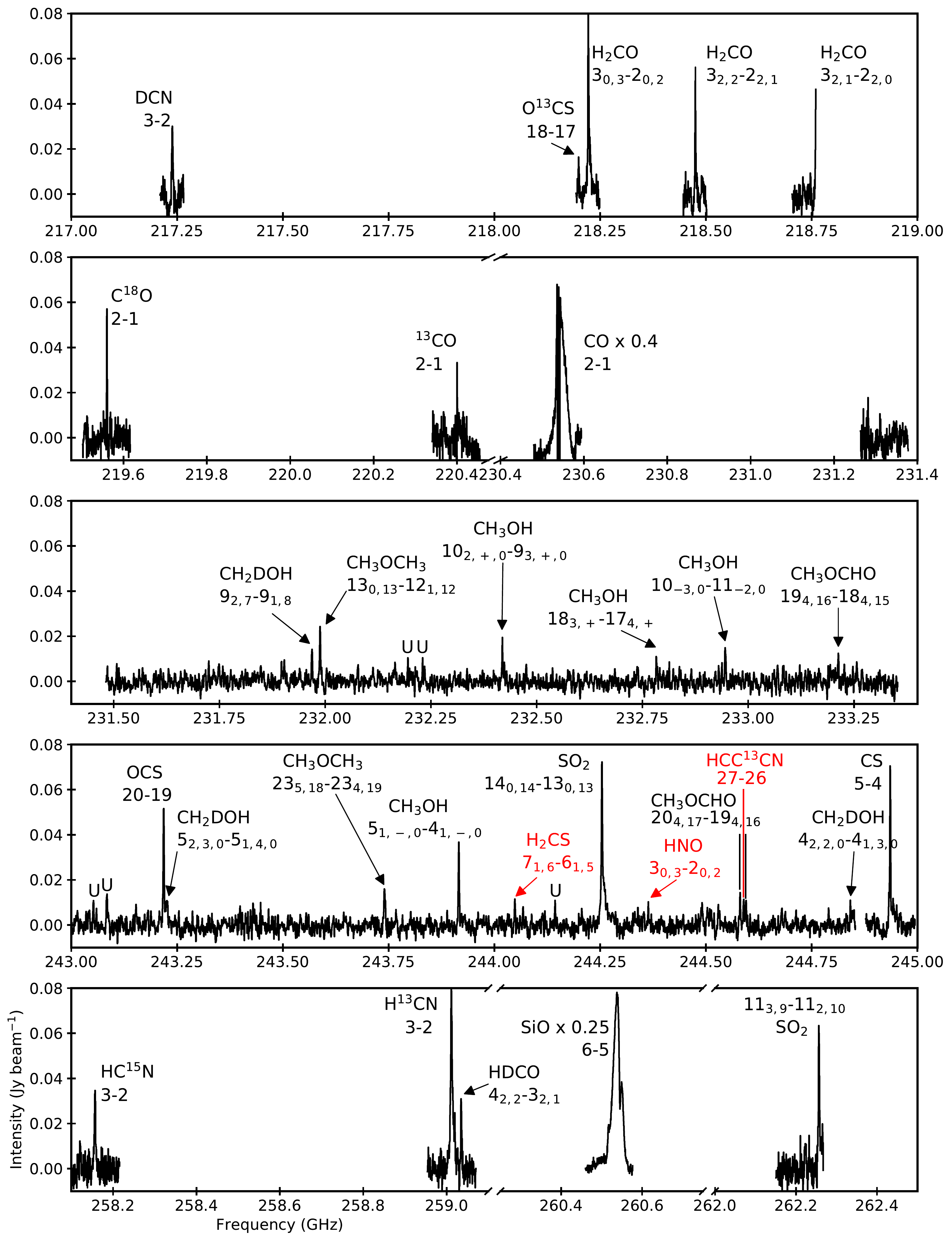}
\caption{Spectrum of the Ser-emb 1 source toward the continuum peak after continuum substraction, re-binned to a spectral resolution of 0.5 MHz. Tentative assignments are indicated in red. Unassigned lines above 4$\sigma$ are marked with a U.}
\label{spec}
\end{figure*}

\begin{deluxetable*}{cccccccccc} 
\tablecaption{Identified molecular transitions toward Ser-emb 1. Tentative assignments are indicated with a $^{*}$. \label{lines1}}
\tablehead{
\colhead{Molecule} & \colhead{Transition} & \colhead{Frequency} & \colhead{log(A$_{ul}$)} & \colhead{$g_{up}$} & \colhead{$E_{up}$} & \colhead{$V_{LSR}$} & \colhead{FWHM} & \colhead{Mom. 0 Int.$^{(a)}$} & \colhead{Mom. 0 rms$^{(a)}$}\\
\colhead{} & \colhead{} & \colhead{(GHz)} & \colhead{(s$^{-1}$)} & \colhead{} & \colhead{(K)} 
& \multicolumn{2}{c}{(km s$^{-1}$)} & \multicolumn{2}{c}{(mJy beam$^{-1}$ km s$^{-1}$)}}
\startdata
CO & 2 $-$ 1 & 230.538 & -6.16 & 5 & 16.60 & \nodata$^{(b)}$ & \nodata$^{(b)}$ & 7719.4 & 277.5$^{(c)}$\\
$^{13}$CO & 2 $-$ 1 & 220.399 & -6.22 & 5 & 15.87 & \nodata$^{(b)}$ & \nodata$^{(b)}$ & 110.8 & 9.8 \\
C$^{18}$O & 2 $-$ 1 & 219.560 & -6.22 & 5 & 15.81 & 8.71 & 2.13 & 118.6 & 6.1 \\
CS & 5 $-$ 4 & 244.936 & -3.52 & 11 & 35.27 & 8.68 & 2.66 & 366.4 & 7.9 \\
SiO & 6 $-$ 5 & 260.518 & -3.04 & 13 & 43.74 & \nodata$^{(b)}$ & \nodata$^{(b)}$ & 16069.6 & 276.2$^{(c)}$\\
DCN & 3 $-$ 2 & 217.238 & -3.49 & 33 & 20.85 & 8.48 & 5.33$^{(d)}$ & 71.3 & 9.3\\ 
H$^{13}$CN & 3$_2$ $-$ 2$_2$ & 259.012 & -3.34 & 33 & 24.86 & 8.95 & 5.51$^{(d)}$ & 317.7 & 7.5 \\
HC$^{15}$N & 3 $-$ 2 & 258.157 & -3.09 & 7 & 24.78 & 8.89 & 3.78& 60.3 & 5.4 \\
HCC$^{13}$CN$^{*}$ & 27 $-$ 26 & 244.589 & -2.94 & 165 & 164.34 & 8.25 & 2.95 & 31.2 & 6.8 \\
HNO$^{*}$ & 3$_{0,3}$ $-$ 2$_{0,2}$ & 244.364 & -4.37 & 38 & 23.46 & 9.01 & 1.49$^{(d)}$ & 26.4 & 7.6 \\
%
N$_2$D$^{+}$ & 3 $-$ 2 & 231.322 & -2.67 & 63 & 22.20 &  \nodata$^{(b)}$ & \nodata$^{(b)}$ & 29.1 & 7.0\\
C$_2$H & 3 $-$ 2 & 262.208 & -5.24 & 7 & 25.16 & \nodata$^{(b)}$ & \nodata$^{(b)}$ & 17.1 & 4.4\\
SO$_2$ & 11$_{3,9}$ $-$ 11$_{2,10}$ & 262.257 & -3.85 &  23 & 82.80 & 8.98 & 3.05 & 173.8 & 7.4  \\
SO$_2$ & 14$_{0,14}$ $-$ 13$_{1,13}$ & 244.254 & -3.79 & 29 & 93.90 & 8.90 & 3.77 & 294.6 & 6.8 \\
OCS & 20 $-$ 19 & 243.218 & -4.38 & 41 & 122.57 & 8.41 & 3.00 & 168.7 & 8.7 \\
O$^{13}$CS & 18 $-$ 17 & 218.199 & -4.52 & 37 & 99.49 & 8.74 & 3.02 &  47.8 & 4.9 \\
H$_2$CS$^{*}$ & 7$_{1,6}$ $-$ 6$_{1,5}$ & 244.048 & -3.68 & 45 & 60.05 & 8.41 &3.37 & 37.7 & 7.6 \\
H$_2$CO & 3$_{0,3}$ $-$ 2$_{0,2}$ & 218.222 & -3.55 & 7 & 20.96 & 8.81 & 3.88 & 214.5 & 5.4 \\ 
H$_2$CO & 3$_{2,2}$ $-$ 2$_{2,1}$ & 218.476 & -3.80 & 7 & 68.09 & 8.76 & 2.87 & 107.5 & 5.2\\ 
HDCO & 4$_{2,2}$ $-$ 3$_{2,1}$ & 259.035 & -3.43 & 9 & 62.86 & 8.75 & 2.56 & 62.7 & 5.1\\
CH$_3$OH & 5$_{1^{-}}$ $-$ 4$_{1^{-}}$ & 243.916 & -4.22 & 11 & 49.66 & 8.64 & 3.33 & 115.6 & 8.5\\
CH$_3$OH & 10$_{2^{+}}$ $-$ 9$_{3^{+}}$ & 232.418 & -4.73 & 21 & 165.40 & 8.45 & 2.77 & 55.2 & 7.4 \\
CH$_3$OH & 10$_{-3}$ $-$ 11$_{-2}$ & 232.946 & -4.67 & 21 & 190.37 & 8.58 &3.56 & 46.5 & 6.5 \\
CH$_3$OH & 18$_{3,+}$ $-$ 17$_{4,+}$ & 232.783 & -4.66 & 37 & 446.53 & 8.70 & 2.77 & 28.31 & 7.2\\
CH$_2$DOH & 4$_{2,2,0}$ $-$ 4$_{1,3,0}$ & 244.841 & -4.32 & 9 & 37.59 & 8.27 & 1.38 & 21.3 & 6.5 \\
CH$_2$DOH & 5$_{2,3,0}$ $-$ 5$_{1,4,0}$ & 243.226 & -4.18 & 11 & 48.40 &  8.94 & 4.21$^{(e)}$ & 86.5$^{(e)}$ & 10.3 \\
CH$_2$DOH & 9$_{2,7,0}$ $-$ 9$_{1,8,0}$ & 231.969 & -4.06 & 19 & 113.11 & 9.38 & 3.62 & 47.2 & 6.5\\
%
CH$_3$OCH$_3$ & 13$_{0,13}$ $-$ 12$_{1,12}$ & 231.988 &  -4.04 & 972 & 80.92 & 8.50 & 3.77$^{(d)}$ & 93.0 & 7.2\\
CH$_3$OCH$_3$ & 23$_{5,18,3}$ $-$ 23$_{4,19,3}$ & 243.739 & -4.10 & 1692 & 287.00 & 8.28 & 4.95$^{(d)}$ & 82.2 & 8.5 \\
CH$_3$OCHO & 19$_{4,16,1}$ $-$ 18$_{4,15,1}$ & 233.213 & -3.74 & 78 & 123.21 & 8.34 & 2.76 & 36.9 & 6.5 \\
CH$_3$OCHO & 20$_{4,17,1}$ $-$ 19$_{4,16,1}$ & 244.580 & -3.68 & 82 & 134.99 & 7.97 & 2.78 & 33.6 & 5.1 \\
CH$_3$OCHO & 20$_{4,17,0}$ $-$ 19$_{4,16,0}$ & 244.594 & -3.68 & 82 & 134.99 & 8.13 & 4.11$^{(e)}$ & 40.1$^{(e)}$ & 7.6 \\
\enddata
\begin{list}{}
	\item $^{(a)}$The moment 0 peak intensities and rms correspond to the individual images cleaned with \textit{robust} = 0.0 and integrated over the full line width (see Sect. \ref{spatial}, and Fig. \ref{mom0}), except for N$_2$D$^{+}$ and C$_2$H, that were imaged with \textit{robust} = 0.5 due to the low S/N of the lines. 
	\item $^{(b)}$ The spectrum of the central pixel could not be fitted to a Gaussian due to the lack of emission at the protostar position (see Fig. \ref{mom0}). 
    \item $^{(c)}$Noise around the region where the emission is located (no line-free channels in the corresponding spectral windows). 
    \item $^{(d)}$Spectrally unresolved multiplets. 
    \item $^{(e)}$Possibly blended with other transitions. 
\end{list}
\end{deluxetable*}

The spectrum at the continuum peak 
(Sect. \ref{obs_spec}) is presented in Fig. \ref{spec}. 
%
Lines were identified 
using the software MADCUBAIJ\footnote{Madrid Data Cube Analysis on ImageJ is a software developed at the Center of Astrobiology (Madrid, INTA-CSIC) to visualize and analyze single spectra and data cubes \citep{rivilla16a,rivilla16b}}, which makes use of the Jet Propulsion Laboratory \citep[JPL;][]{pick98} and the Cologne Database for Molecular Spectroscopy  \citep[CDMS;][]{mull05} spectral catalogs. 
We attempted to assign every observed line with an intensity above 4$\sigma$ (the rms of the extracted spectra roughly corresponds to the Channel rms in Table \ref{spw}) to a particular molecular transition, starting with the strong transitions which could unambiguously be assigned to small, abundant molecules. For weaker lines, we generally only found one plausible molecular candidate and tentatively assigned it. Then,   
for every putative assignment, we checked for competing identifications, and looked for all the detectable transitions of that particular molecule in our spectral range, ruling out that there were any missing lines in our data. 

Of the 36 observed lines that fulfilled the above criteria, those molecules presenting at least two $>$4$\sigma$ lines 
(H$_2$CO, CH$_3$OH, CH$_2$DOH, CH$_3$OCH$_3$, CH$_3$OCHO, and SO$_2$), 
lines from more than one isotopolog (CO, $^{13}$CO, and C$^{18}$O; HDCO; DCN, H$^{13}$CN, and HC$^{15}$N; O$^{13}$CS and OCS), or one line intense enough ($>$10$\sigma$) to be unambiguously assigned (CS, SiO) are considered detections. 
Species with just one detected line below 10$\sigma$ are considered tentative detections (H$_2$CS, HNO, HC$^{13}$CCN), and are indicated in red in Fig. \ref{spec}. 
Only five lines with an intensity above 4$\sigma$ remained unassigned at 232.195, 232.230, 243.055, 243.085, and 244.145 GHz (marked with a U in Fig. \ref{spec}). 

A list of the identified molecular transitions in  Fig. \ref{spec} 
(as well as the N$_2$D$^+$ 3 $-$ 2 and C$_2$H 3 $-$ 2  lines, detected away from the continuum peak) is presented in Table \ref{lines1}, arranged by chemical families in ascending order of complexity. 
The line parameters were extracted from the JPL catalog, except for CH$_3$OCH$_3$, whose line parameters were extracted from the CDMS catalog. 
In particular, the values for CH$_3$OH, CH$_2$DOH, CH$_3$OCH$_3$, and CH$_3$OCHO used in Sect. \ref{sec:rot} are extracted from \citet{xu08}, \citet{pearson12}, \citet{endres09}, and \citet{ilyushin09}, respectively.
Among the detected molecules we note two H$_2$CO lines and four CH$_3$OH lines, as well as additional lines from the single-deuterated isotopologs HDCO (one line) and CH$_2$DOH (three lines). 
Formaldehyde and methanol are considered COM precursors, and often refered to as 0th generation COMs \citep[see, e.g.,][]{oberg09,herbst09}. 
Two COMs were also detected at the continuum peak position of Ser-emb 1: dimethyl eter (CH$_3$OCH$_3$) and methyl formate (CH$_3$OCHO). 
In particular, we detected two unresolved multiplets of CH$_3$OCH$_3$ with signal-to-noise ratios of 7 and 12, and three CH$_3$OCHO lines with signal-to-noise ratios of 5$-$6. 

\subsubsection{Rotational temperature and column densities\footnote{We note that the values presented in this Section are consistent, within errors, with those reported in \citet{jenny19}. The differences are due to different fitting techniques used in both cases.}}\label{sec:rot}

\begin{figure*}
\centering
\includegraphics[width=17cm]{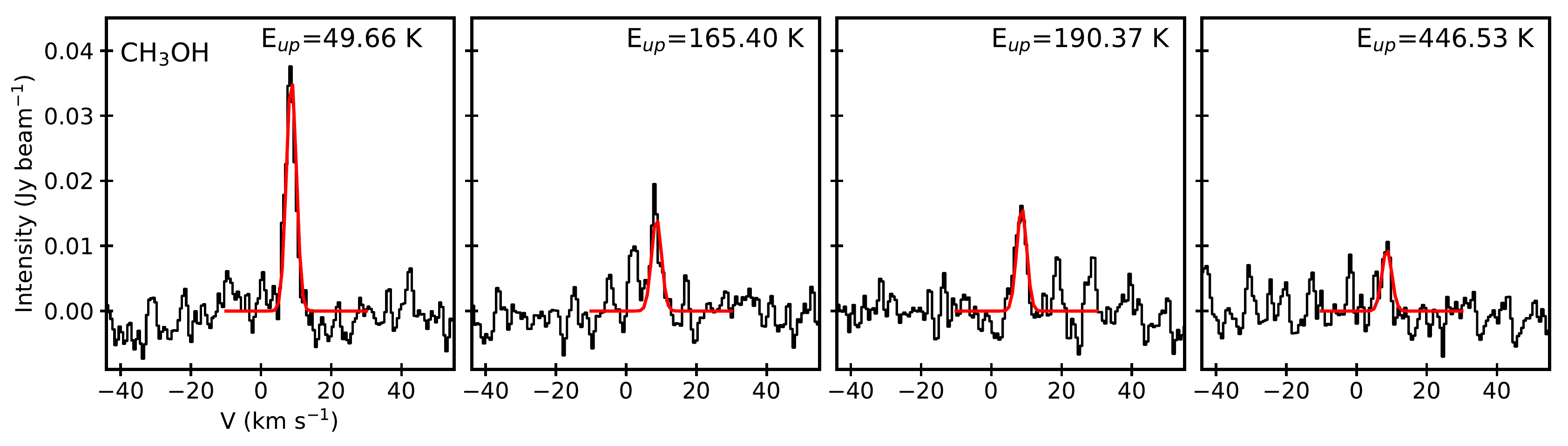}
\caption{Gaussian fits in red to the observed CH$_3$OH lines in black (see Table \ref{lines1}) used to calculate the integrated flux density for Eq. \ref{nueq}}
\label{ch3ohlines}
\end{figure*}

\begin{figure}
\plotone{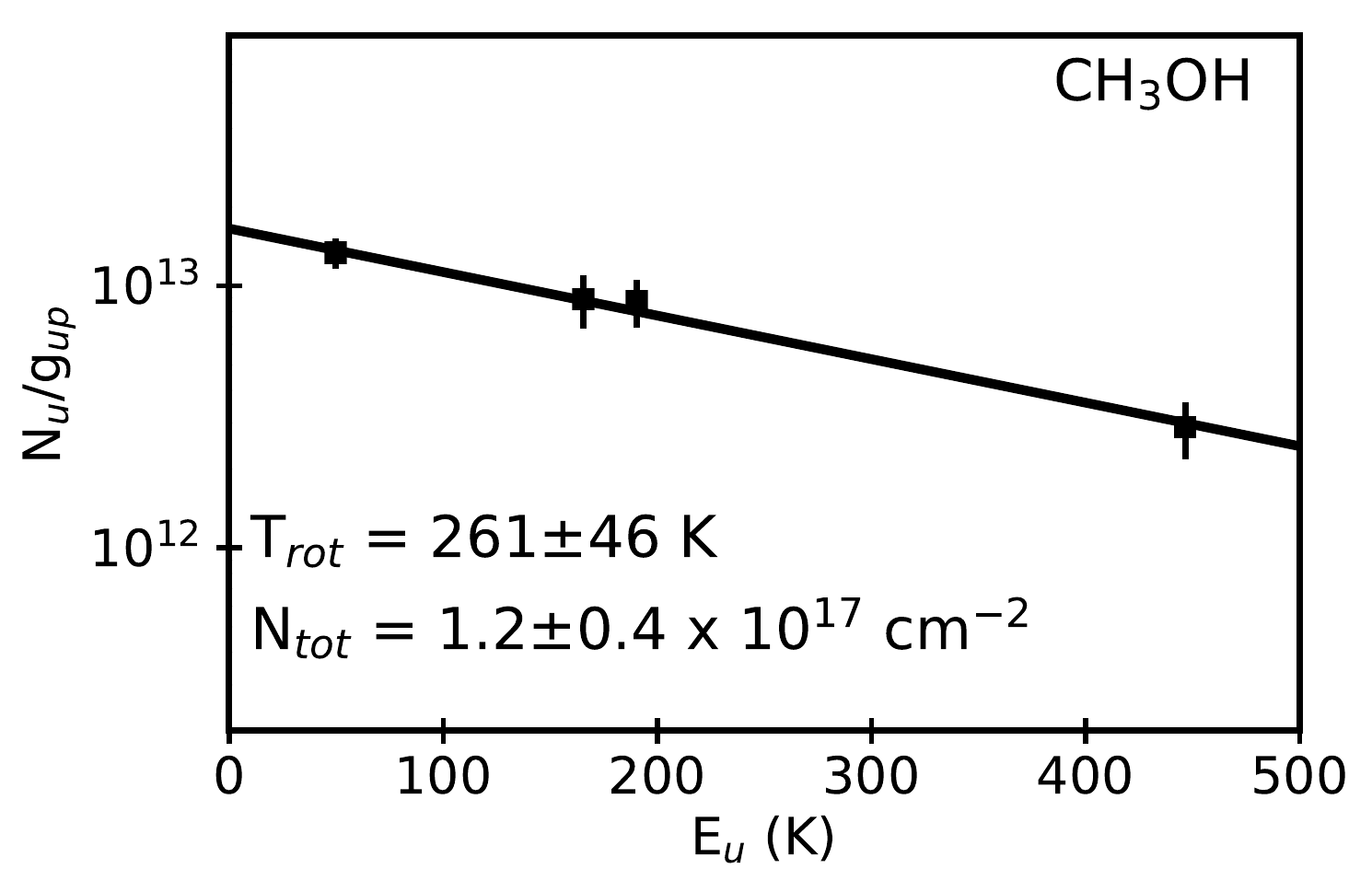}
\caption{Rotational diagram for CH$_3$OH. The three squares correspond to the three detected transitions in our dataset (see Fig. \ref{ch3ohlines}). The plotted 1$\sigma$ errors include an additional 10\% to the integrated flux density errors provided by the Gaussian fit to account for calibration errors, since we are using lines observed with two different frequency settings. The best fit to the data is represented by a solid line, and the derived rotational temperature and column density are indicated in the plot.}
\label{ch3ohrot}
\end{figure}

\begin{figure*}
\centering
\includegraphics[width=17cm]{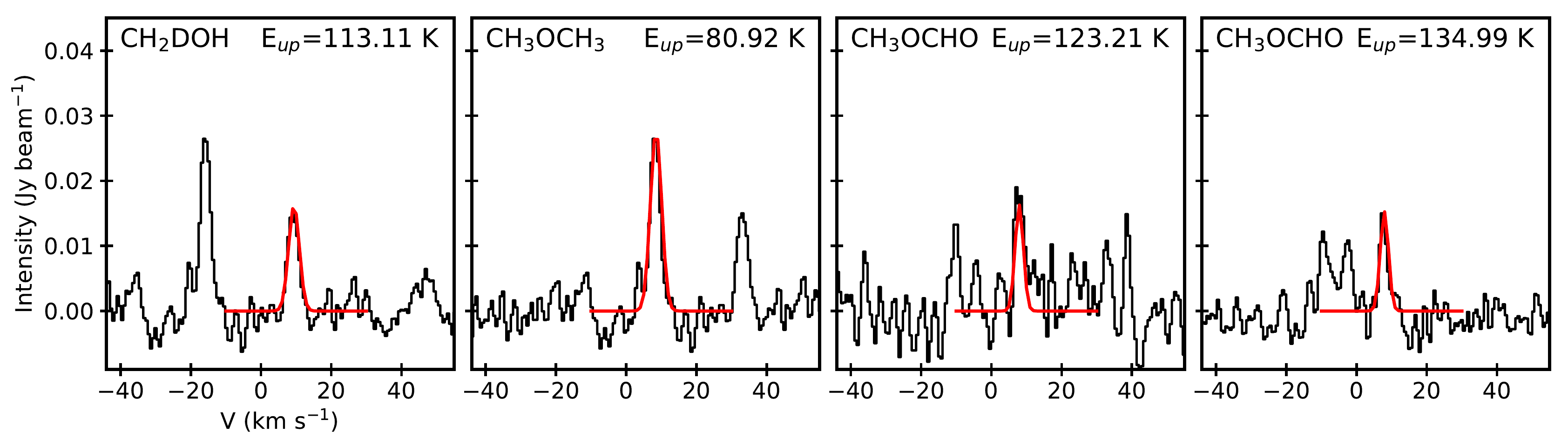}
\caption{Gaussian fits in red to the observed CH$_2$DOH, and CH$_3$OCH$_3$, and CH$_3$OCHO lines (see Table \ref{lines1}) used to calculate the integrated flux density for Eq. \ref{nueq}. 
\label{otherlines1}}
\end{figure*}

For those species with more than one detected transition, the rotational diagram can be used to derive excitation temperatures and column densities \citep[, see Sect. \ref{app-rot}]{goldsmith99}. 
Figure \ref{ch3ohrot} shows the rotational diagram for CH$_3$OH using the spectrum extracted at the continuum peak. 
We attempted to perform similar calculations for other molecules, but a combination of low S/N and/or a lack of lines spanning a wide range of E$_{up}$ resulted in unconstrained excitation temperatures and column densities. 

To extract the CH$_3$OH line fluxes and flux uncertainties, we fitted a Gaussian to each observed CH$_3$OH line (see Fig. \ref{ch3ohlines}) using the  
Python function \textit{curve\_fit} with a fixed $V_{LSR}$ = 8.64 km s$^{-1}$, and FWHM = 3.33 km s$^{-1}$, corresponding to the position and width of the brightest methanol line detected. 
The integrated flux densities calculated from the Gaussian fits are roughly equal to the moment 0 peak intensities listed in Table \ref{lines1}.
In Ser-emb 1, the methanol emission originates in a compact region around the protostar (see Sect. \ref{spatial}). 
For the images cleaned with a robustness parameter of 0.5 (where the spectrum has been extracted from), the size of the beam is $\Omega_{beam}$ = 0.55$\arcsec$ $\times$ 0.45$\arcsec$ (Table \ref{spw}), while the size of the methanol emitting region is $\Omega_{source}$ $\le$ (0.29 $\pm$ 0.15)$\arcsec$ $\times$ (0.12 $\pm$ 0.10)$\arcsec$. 
The source size was estimated with a 2D Gaussian fit of the 10$_{-3}$ $-$ 11$_{-2}$\footnote{This transition, with an upper level energy E$_{up}$ = 190 K, is a better tracer of the compact methanol emitting region (a hot corino, see Sect. \ref{spatial}) than the brightest  5$_{1^{-}}$ $-$ 4$_{1^{-}}$ line, since the latter has an upper level energy E$_{up}$ = 50 K, and could be tracing cooler, more extended material.} moment 0 map  after deconvolving the beam size.  
However, this estimated source size is the largest source size consistent with our observations, and the actual size of the methanol emitting region could be smaller.  
For the rotational diagram, we corrected the observed flux densities from beam dilution using the minimum filling factor of $\sim$7.5 in Eq. \ref{nueq} that corresponds to the maximum source size. Therefore, the calculated column density should be considered a lower limit.

The rotational diagram fit was performed with the function \textit{curve\_fit} in Python. 
The derived column density for CH$_3$OH is $N_T$ = (1.2 $\pm$ 0.4) $\times$ 10$^{17}$ cm$^{-2}$ (corresponding to the largest source size consistent with our observations), and the rotational temperature is $T_{rot}$ = 261 $\pm$ 46 K  
(i.e., $T_{rot}>$100 K, above the water ice desorption temperature).  
The optical thickness of the brightest detected line is $\tau$ = 0.1. 
Therefore, we expect the optically thin line assumption (Sect. \ref{app-rot}) to be valid for all observed transitions.
We have also fitted the methanol rotational diagram excluding the brightest line (E$_{up}$ = 49.66 K), and get the same column density and a rotational temperature. 
This suggests that the optical thickness of the 5$_{1-}$ $-$ 4$_{1-}$ line is not impacting the calculated column density.
However, we note that these values should be considered lower limits, as stated above, since the actual size of the methanol emitting region could be smaller, increasing the column density up to a factor of 3 \citep{jenny19}. 
We therefore cannot completely exclude that the 5$_{1-}$ $-$ 4$_{1-}$ line is optically thick. 
Observations of more lines of CH$_3$OH and its isotopologs are needed to better constrain the methanol column density and address the optical depth issues.

We adopted the rotational temperature derived for CH$_3$OH to estimate the column densities of 
CH$_2$DOH and the two detected COMs,  
applying Eq. \ref{rot} to their brightest detected transition (see Fig. Fig. \ref{otherlines1}), 
and the same minimum filling factor as for CH$_3$OH, since we expect these species to emit from the same region (see Sect. \ref{spatial}). 
The results for CH$_2$DOH (using the 9$_{2,7,0}$ $-$ 9$_{2,7,0}$ transition), 
CH$_3$OCH$_3$ (treating the unresolved multiplet 13$_{0,13}$ $-$ 12$_{1,12}$ as a single line), 
and CH$_3$OCHO (averaging the total column density calculated from the two unblended transitions, since the strongest transition is blended) are presented in Table \ref{abundances}. 
We have corrected the CH$_3$OCHO column density by a factor of 2.1 (corresponding to a rotational temperature of 261 K) to take into account the higher torsional states that are not included in the partition function extracted from the JPL catalog \citep{favre14}.
The estimated column densities are (4.8 $\pm$ 1.8) $\times$ 10$^{16}$, (9.2 $\pm$ 3.8) $\times$ 10$^{16}$, and (9.1 $\pm$ 3.6) $\times$ 10$^{16}$, for CH$_2$DOH, CH$_3$OCH$_3$, and CH$_3$OCHO, respectively. Again, these values should be considered lower limits since we have used the minimum filling factor. 
These column densities imply a D/H ratio for methanol of 0.40, and CH$_3$OCH$_3$/CH$_3$OH and CH$_3$OCHO/CH$_3$OH ratios of 0.77 and 0.76, respectively.  
These ratios are high, 
and could be affected by the uncertainties in the estimation of the methanol column density explained above. 
If we instead used, for example, the D/H ratio recently reported for methanol in HH 212 \citep[0.27,][]{lee17}, the CH$_3$OH column density would be 1.8 $\times$ 10$^{17}$; while applying the D/H ratio found in B1b-S for methyl formate \citep[0.02,][]{marcelino18} to the CH$_2$DOH column density leads to a CH$_3$OH column density one order of magnitude higher (2.4 $\times$ 10$^{18}$). 
In addition, we note that methanol emission often presents an extended component around protostars \citep[see, e.g.,][]{oberg13}, that could be filtered out in these kind of observations that lack short baselines, also leading to an underestimation of its column density. 
Therefore 
the reported ratios should be treated as upper limits.

\begin{deluxetable}{ccc}
\caption{COMs column densities in the hot corino of Ser-emb 1.\label{abundances}}
\tablehead{
\colhead{Molecule} & \colhead{$N_T$} & \colhead{$N_T$/$N_T$(CH$_3$OH)}\\
& \colhead{(molecules cm$^{-2}$)} &}
\startdata
CH$_3$OH & $\ge$(1.2 $\pm$ 0.4)\ $\times$ 10$^{17}$ & 1.00 \\ 
CH$_2$DOH & $\ge$(4.8 $\pm$ 1.8) $\times$ 10$^{16}$ & 0.40\\
CH$_3$OCH$_3$ & $\ge$(9.2 $\pm$ 3.8) $\times$ 10$^{16}$ & 0.77\\
CH$_3$OCHO & $\ge$(9.1 $\pm$ 3.6) $\times$ 10$^{16}$ & 0.76\\
\enddata
\tablecomments{These column densities are calculated assuming a maximum source size of $\Omega_{source}$ = (0.29 $\pm$ 0.15)$\arcsec$ $\times$ (0.12 $\pm$ 0.10)$\arcsec$, as extracted from a 2D Gaussian fit after deconvolving the beam size, and should therefore be considered lower limits (see text).}
\end{deluxetable}

\subsection{Spatial distributions}\label{spatial}

\begin{figure*}
\centering
\includegraphics[width=17cm]{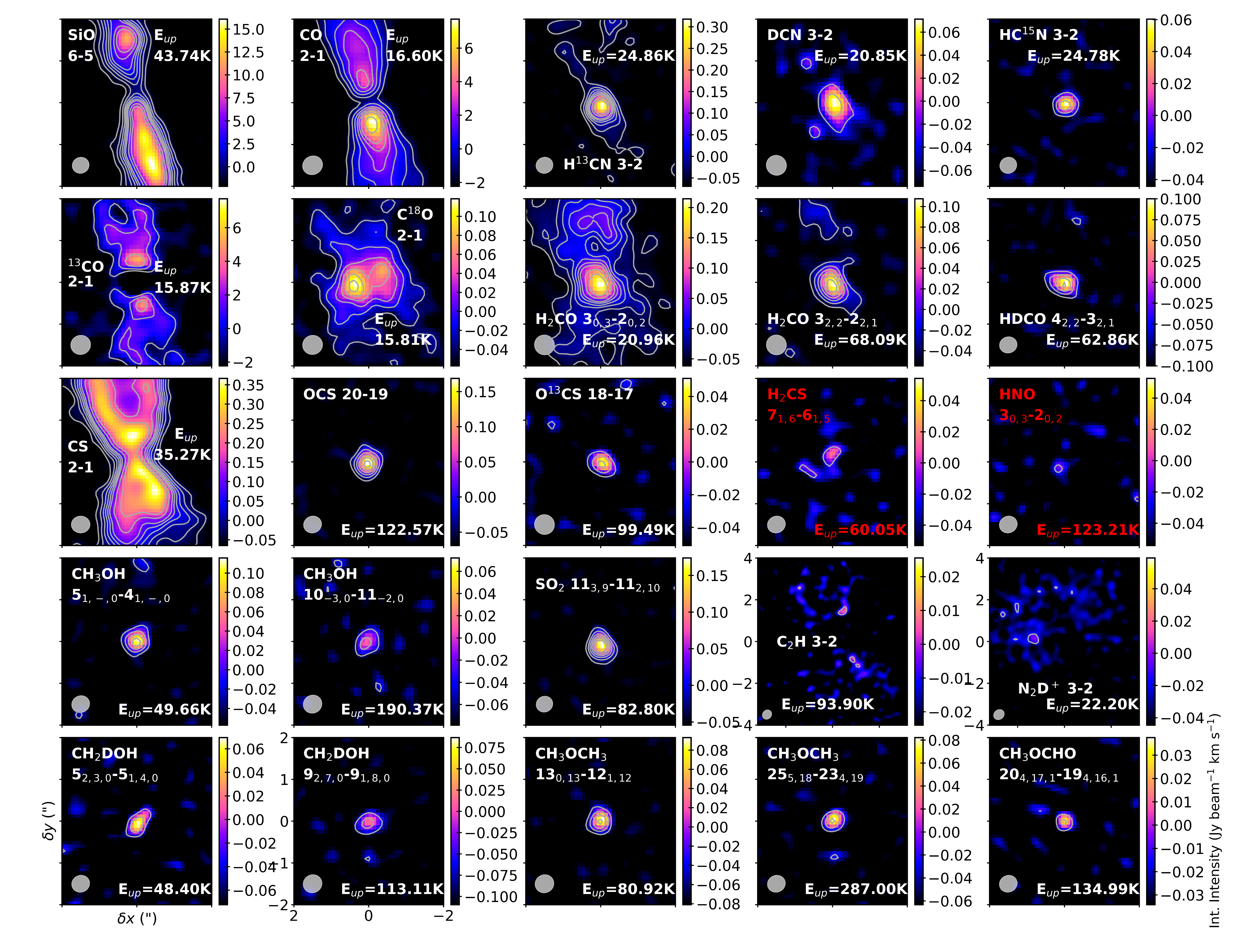}
\caption{Integrated intensity (moment 0) maps, along with the 3, 6, 9, 12, 15, 18, 24, and 30$\sigma$ contours of some transitions identified toward the continuum peak of Ser-emb 1 (maps are clipped at 1$\sigma$). The rms for the contours is indicated in Table \ref{lines1} for each line. 
The size of the synthesized beam ($\sim$ 0.50$\arcsec$ $\times$ 0.42$\arcsec$) is shown in the lower left of every panel (see also Sect. \ref{app-obs}).  
\label{mom0}}
\end{figure*}

To explore the spatial distributions of the detected molecules, the individual data cubes cleaned with a robustness parameter of 0.0 for a better angular resolution (Sect. \ref{obs_im}) were integrated over the full line width of the emission lines. 
Fig. \ref{mom0} shows the resulting integrated emission images (moment 0 maps) of all the molecules listed in Table \ref{lines1}. For molecules where multiple lines were detected with substantially different upper level energies, two lines are shown. The integrated intensity peaks and the moment 0 maps rms\footnote{The moment 0 map rms of every transition in Table \ref{lines1} is calculated at the location of the emission using a moment 0 map with the same number of line-free channels, except for CO and SiO, whose data cubes did not have enough line-free channels. For those cases, the rms is averaged from emission-free regions in the line moment 0 maps.} are listed in Table \ref{lines1}.

Among the small species (4 atoms or less) most, but not all, present emission extended over the 
the slow outflow region traced by the low-velocity CO 2 $-$ 1 emission (see Sect. \ref{overview}), up to $\sim$4$\arcsec$ or $\sim$1600 au of the central protostar in the case of CS or H$_2$CO. 
The exceptions are, on one hand, N$_2$D$^+$, whose emission is extended to the outer cold envelope where CO molecules are frozen onto the dust grains \citep[see, e.g., ][]{bergin01}, with no emission detected within $\sim$1$\arcsec$ or $\sim$400 au from the central protostar; and on the other hand, OCS and SO$_2$ that present only compact, barely resolved emission. 
By contrast, the emission of all organic molecules with 5 or more atoms is unresolved in a compact region around the protostar.  
CH$_3$OH and CH$_2$DOH are detected within $\sim$0.4$\arcsec$ or $\sim$160 au from the central protostar. 
The size of the emitting region for the 5$_{1,-,0}$ $-$ 4$_{1,-,0}$ methanol line, obtained with a 2D Gaussian fit after deconvolving the beam size, is (0.26 $\pm$ 0.08)$\arcsec$ x (0.12 $\pm$ 0.11)$\arcsec$ (this is actually the maximum source size consistent with the data, see Sect. \ref{sec:rot}). Methanol would be thus emitting within a radius of $\sim$50 au from the central protostar. 
The emission from CH$_3$OCH$_3$ and CH$_3$OCHO is contained within the beam size 
and it could not be deconvolved from the beam. 
The detection of these organic molecules in a compact region around the protostar is not only a matter of sensitivity, 
since the peak integrated intensity of some lines presenting only compact emission (e.g., CH$_3$OH 5$_{1,-,0}$ $-$ 4$_{1,-,0}$) is higher than that of some lines showing extended emission (e.g., H$_2$CO 3$_{2,2}$ $-$ 2$_{2,1}$, see Table \ref{lines1}). 
Therefore, CH$_3$OH and the two COMs CH$_3$OCH$_3$ and CH$_3$OCHO are present in a compact (the beam radius is $\sim$100 au, see Sect. \ref{obs_im}), high-temperature region around the central protostar.  
This is consistent with the high excitation temperature of CH$_3$OH derived in the previous section.

\subsection{Rotation signature\label{disk}}

\begin{figure}
\centering
\includegraphics[width=8.5cm]{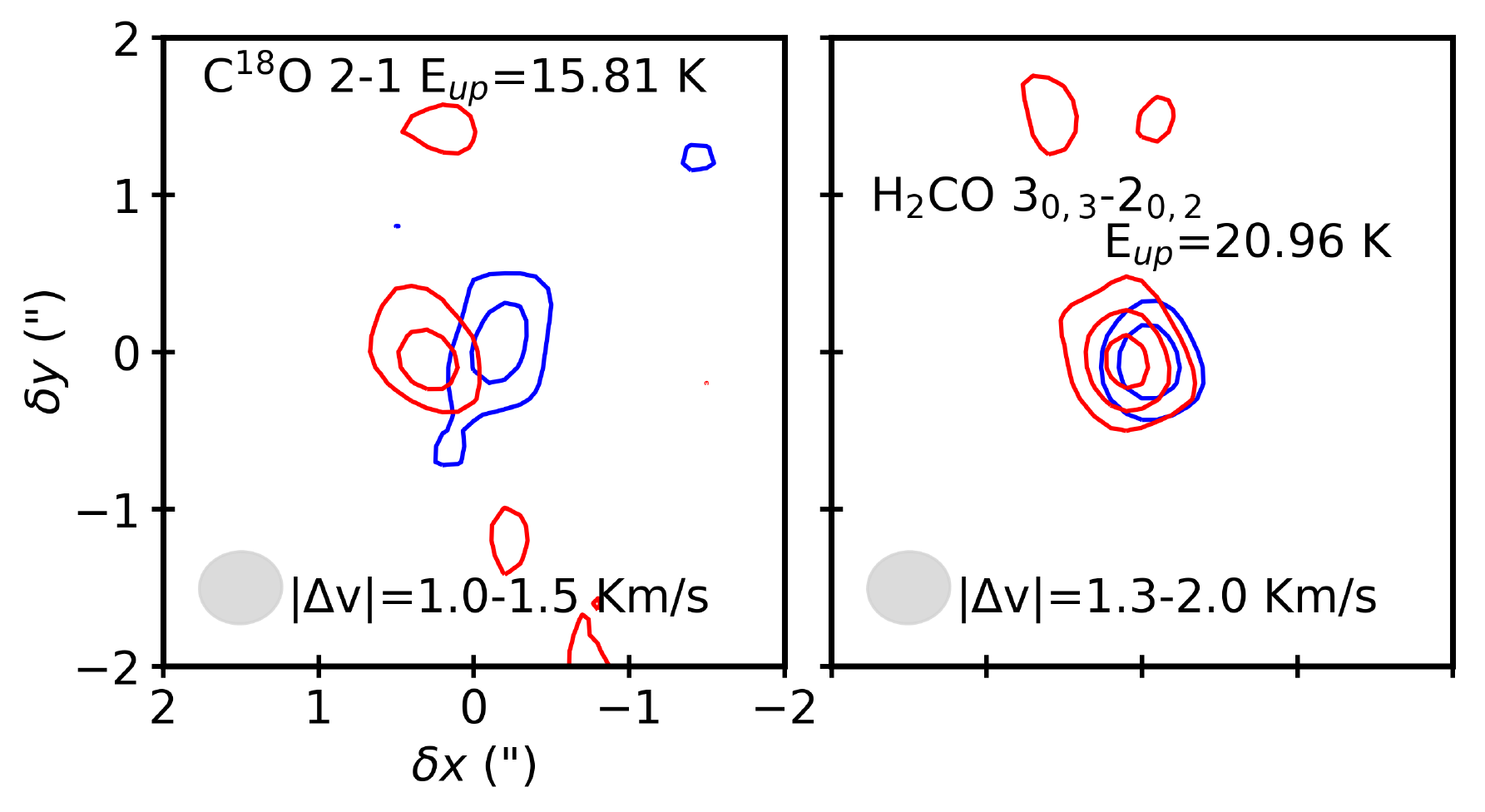}
\caption{3, 6, and 9$\sigma$ contours of the high-velocity blue- and red-shifted moment 0 maps corresponding to different molecular transitions showing spatially resolved blue- and red-shifted emission. The range of velocities with respect to $V_{LSR}$ is indicated in every panel. The size of the synthesized beam ($\sim$ 0.50$\arcsec$ $\times$ 0.42$\arcsec$) is shown in the lower left of every panel (see also Sect. \ref{app-obs}). 
\label{diskfig}}
\end{figure}

In Fig. \ref{overview2}, we show 
that the high-velocity C$^{18}$O 2 $-$ 1 emission is compact, and spatially distributed from red- to blue-shifted velocities orthogonally to the direction of the outflow, which could be an indication of rotation of this molecule around the central object. 
We similarly plotted all detected lines and found 
a comparable velocity gradient 
in the H$_2$CO 3$_{0,3}$ $-$ 2$_{0,2}$ 
high-velocity map (Fig. \ref{diskfig}). 
%
We note that the emission originating from CH$_3$OH and the CH$_3$OCH$_3$ and CH$_3$OCHO COMs appears unresolved, and a similar velocity pattern on smaller scales cannot be excluded. 

A priori, a velocity gradient across a protostar can have several different origins: outflows (see Fig. \ref{overview2}), infall, and rotation. In the latter case, it is not given that rotation automatically implies the presence of a Keplerian (rotationally supported) disk around the central protostar, since it also trace the rotating motion of the infalling envelope. %
In Ser-emb 1, several aspects of the observed emission may point to 
the former 
as the origin of the velocity gradient. 
First, there is no sign of protostellar multiplicity and thus reason to expect a second outflow associated with this source
Second, the velocity gradient is orthogonal to the observed outflow, which is assumed to be the axis of rotation of a potential protostellar disk.  
Third, the high-velocity emission showing the velocity gradient is very compact, i.e. only visible on scales associated with protostellar disks ($<$200 au radius).  
Still we cannot exclude from the maps alone that we are observing an infalling-rotating envelope rather than a rotationally supported disk \citep[see, e.g.,][]{tobin12a,oya16,oya17,jacobsen18}. 

\begin{figure}
\centering
\includegraphics[width=8.5cm]{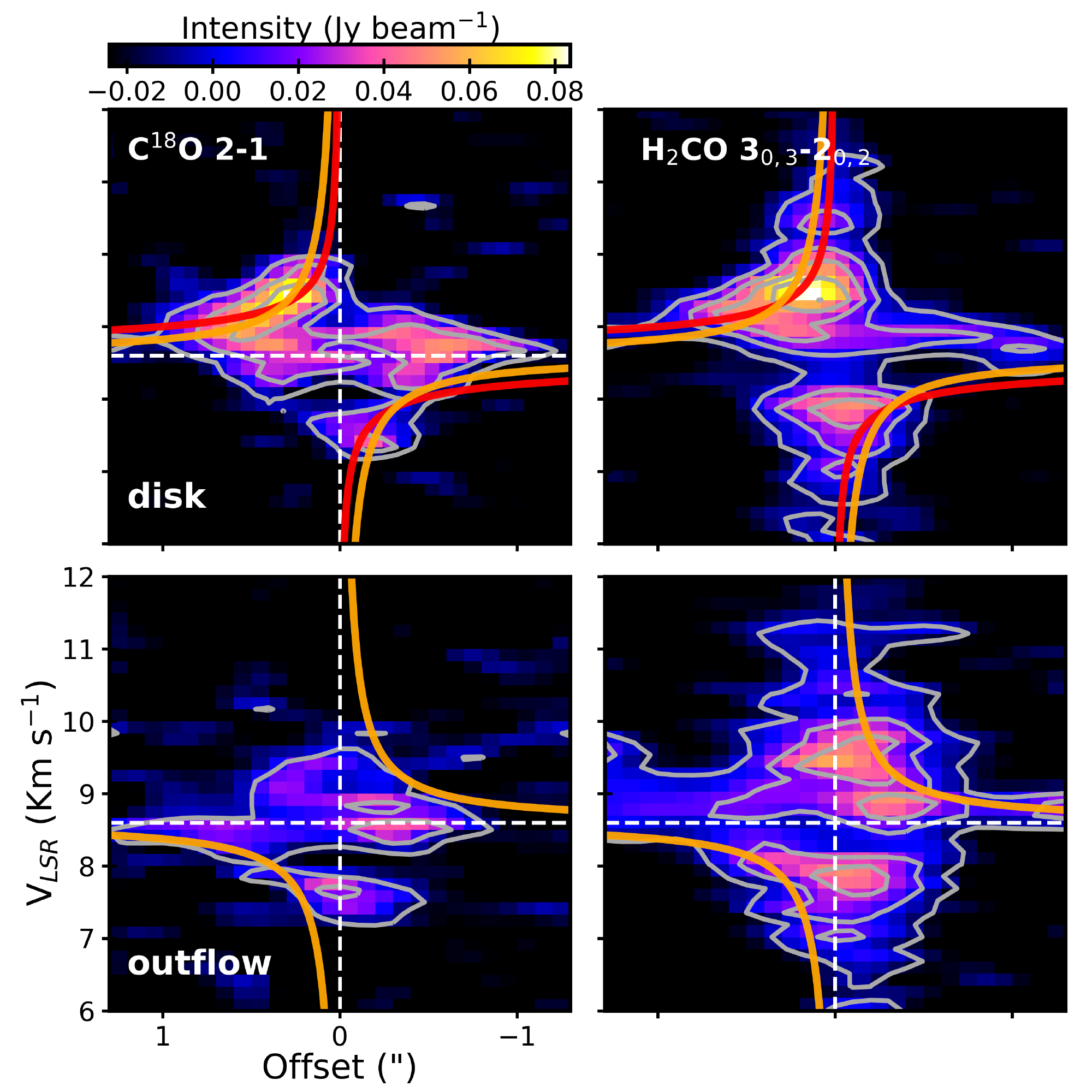}
\caption{PV diagrams of a 0.24$\arcsec$ width slice along the direction perpendicular to the outflow (PA = -77.7$^{\circ}$, top panels) and the direction of the outflow, assumed to be the axis of rotation (PA = 12.2$^{\circ}$, bottom panels) at both sides of the continuum peak position (offset = 0$^{\prime\prime}$) corresponding to the molecular lines in Fig. \ref{diskfig}. The position of the central protostar and the radial velocity of the source are indicated with dashed, white lines. The 3, 6, 9, 12, 15$\sigma$ contours are overlaid in grey (see Table \ref{spw} for the rms corresponding to each transition). 
Pure Keplerian ($v \propto R^{-0.5}$, red) and infall ($v \propto R^{-1}$, orange) curves are shown, when appropriate. 
\label{pv}}
\end{figure}

In order to further study these rotation signatures, the top panels of Fig. \ref{pv} show the position-velocity (PV) diagrams along the direction perpendicular to the outflow, for the two transitions in Fig.  \ref{diskfig}. 
The C$^{18}$O 2 $-$ 1, and H$_2$CO 3$_{0,3}$ $-$ 2$_{0,2}$ lines show a spin-up feature toward the protostar. 
The spin-up feature consists in a velocity gradient from one side of the outflow axis to the other, 
with the red-shifted emission ($V_{LSR}$ $>$ 8.6 km s$^{-1}$, above the horizontal dashed line) mainly located to the east of the central protostar position (offset $>$ 0$^{\prime\prime}$, to the left of the vertical dashed line), 
and the blue-shifted emission ($V_{LSR}$ $<$ 8.6 km s$^{-1}$, below the horizontal dashed line) preferentially detected to the west (offset $<$ 0$^{\prime\prime}$, to the right of the vertical dashed line);  
and with 
higher 
velocities closer to the protostar. 
A similar feature in the PV diagram was also observed in \citet{lee17} for the red- and blue-shifted emission of CH$_3$OH and CH$_2$DOH, that could be reproduced with a Keplerian rotation model, thus indicating that these species trace the disk atmosphere in HH 212. In our case,  
a Keplerian rotation curve ($v \propto R^{-0.5}$, shown in red in Fig. \ref{pv}) is able to reproduce the observed spin-up feature, but the model considering an infalling-rotating envelope ($v \propto R^{-1}$, shown in orange in Fig. \ref{pv}) cannot be ruled out.  
On the other hand, the spin-up feature is not clearly observed in the PV diagrams along the direction of the outflow (bottom panels of Fig. \ref{pv}), where red- and blue-shifted emission is more symmetrically detected to both the north (offset $>$ 0$^{\prime\prime}$) and the south (offset $<$ 0$^{\prime\prime}$) of the central protostar position. 
This cannot be reproduced by the same infalling-rotating envelope model ($v \propto R^{-1}$, shown in grey in Fig. \ref{pv}) that would locate the red-shifted emission preferentially to the south of the source, and the blue-shifted emission to the north. 
However, these PV diagrams show non-negligible infall velocities along the minor axis of the potential protostellar disk (especially in the case of H$_2$CO), unlike methanol and deuterated methanol in \citet{lee17}.
%

In summary, we have detected a rotation signature orthogonal to the outflow in two molecular lines, which 
are consistent Keplerian rotation, 
but observations with higher spatial resolution (and a higher signal-to-noise ratio) are needed to unambiguously distinguish from the rotating motion of the infalling envelope. 

\section{Discussion\label{discussion}}

\subsection{Hot corino chemistry in Ser-emb 1}

According to the definition in \citet{herbst09}, a warm inner envelope passively heated by the central protostar to typical temperatures around 100 K or warmer, with typical sizes of 100 au or less, associated with complex organic molecule emission is considered a hot corino. 
The presence of CH$_3$OH, CH$_3$OCH$_3$, and CH$_3$OCHO 
emitting in a compact region (beam radius $\sim$100 au, Sect. \ref{obs_im}) around the central protostar in Ser-emb 1 (Sect. \ref{spatial}), with a measured rotational temperature for CH$_3$OH of $T_{rot}$ = 249 $\pm$ 42 K (Sect. \ref{sec:rot}) 
means that Ser-emb 1 fulfills the above criteria, with the 
addition that the hot corino could be associated to an incipient protostellar disk (see Sect. \ref{disk}), as it is the case in, e.g., HH 212 \citep{lee17,codella18,lee19}. 

As stated in Sect. \ref{sec:intro}, not all warm compact cores present a hot core/corino chemistry. Some sources show a different chemistry characterized by unsaturated carbon chains, and are usually known as warm carbon chain chemistry (WCCC) sources \citep[see, e.g.,][]{lefloch18}
While our observations do not constitute an unbiased survey, the non-detection of C$_2$H toward the central protostar position in Ser-emb 1 (Sect. \ref{hotcorino}) implies a hydrocarbon-poor, organic chemistry  similar to other detected hot corinos.

\begin{deluxetable*}{ccccccc}
\caption{COMs column densities in Class 0 hot corinos. Sources with only single-dish observations are marked with a $^{*}$. \label{comcomparison}}
\tablehead{\colhead{Source} & \multicolumn{3}{c}{$N$ ($\times$ 10$^{16}$ cm$^{-2}$)} & \colhead{Reference} & \colhead{L$_\odot$} & \colhead{Reference} \\
& CH$_3$OH & CH$_3$OCH$_3$ & CH$_3$OCHO}
\startdata
IRAS 2A & 500$^{+290}_{-180}$ & 5.0$^{+2.9}_{-1.0}$ & 7.9$^{+4.6}_{-1.6}$ & \citet{taquet15} & 36 & \citet{karska13}\\
IRAS 16293B & 1000 & 24 & 26 & \citet{jorgensen18}$^{a}$ & 22 & \citet{crimier10}\\
L483 & 1700 & 8.0 & 13 & \citet{jacobsen18} & 10-14 & \citet{jacobsen18}\\
IRAS 4A2 & \nodata & 4.5$^{+0.3}_{-0.2}$ & 3.5$^{+0.2}_{-0.4}$ & \citet{lopezsepulcre17} & 9.1 & \citet{karska13}\\
HH 212 & 61.0$\pm$33.0 & \nodata & 3.2$\pm$1.2 & \citet{lee19} & 9 & \citet{zinnecker92}\\
IRAS 4B$^{*}$ & 34.2$^b$ & $<$6.5$^c$ & 8.9$\pm$6.5 & \citet{bottinelli07} & 6 & \citet{jorgensen02}\\
Ser-emb 1$^d$ & 12$\pm$4 & 9.2$\pm$3.8 & 9.1$\pm$3.6 & This work & 4.1 & \citet{enoch11}\\
B335 & \nodata & 0.19$\pm$0.02 & 2.6$\pm$0.3 & \citet{imai16} & 0.72 & \citet{evans15}\\
B1b-S & 35 & 1.0 & 1.5 & \citet{marcelino18} & 0.49 & \citet{pezzuto12}\\
\enddata
\tablecomments{
$^{a}$At the position offset by 0.5$\arcsec$ from the continuum peak. Uncertainties are estimated to be about 20\%.
$^b$From the CH$_3$OCHO/CH$_3$OH ratios in \citet{herbst09}.
$^c$From the CH$_3$OCH$_3$/CH$_3$OH ratio in \citet{herbst09}.
$^d$Assuming a source size of $\Omega_{source}$ = (0.29 $\pm$ 0.15)$\arcsec$ $\times$ (0.12 $\pm$ 0.10)$\arcsec$ (see Sect. \ref{sec:rot}).}
\end{deluxetable*}


Table \ref{comcomparison} compares the estimated column densities of CH$_3$OH and the two COMs reported in this paper  (CH$_3$OCH$_3$ and CH$_3$OCHO) in Ser-emb 1 with those reported in the literature for the other eight Class 0 hot corinos. The bolometric luminosities of the sources are also indicated. 
Roughly, the most luminous sources (IRAS 2A, IRAS 16293B, and L483) present methanol column densities on the order of $\sim$10$^{18-19}$, while the column densities estimated for the rest of hot corinos are on the order of $\sim$10$^{17}$, cosnsitent with the value presented here for Ser-emb 1. 
On the other hand, the column densities of dimethyl ether and methyl formate present a larger scatter across the sample of Class 0 hot corinos, with no obvious trend. 
The COM abundances with respect to CH$_3$OH across the sample of detected hot corinos  (and also the different evolutionary stages during star formation) are further discussed in \citet{jenny19}.

\subsection{A protostellar disk in Ser-emb 1?}\label{diskdisc}

\citet{enoch11} already suggested the presence of a partially resolved disk structure around the central protostar in Ser-emb 1 due to the substantial $\sim$230 GHz continuum flux detected at intermediate \textit{uv} distances (30$-$100 k$\lambda$). 
This corresponds to a disk radius above 170 au, of the same order as other confirmed protostellar disks in Class 0 sources \citep[50$-$150 au,][]{choi10,tobin12,murillo13,yen13,codella14,lindberg14,oya16,yen17,lee17}. 
The value reported in \citet{enoch11} agrees well with our estimated radius of $\sim$200 au for the C$^{18}$O high-velocity emission 
in Fig. \ref{diskfig}. The disk mass estimated from the continuum flux at 50 k$\lambda$ is (0.28 $\pm$ 0.14) $M_{\odot}$ \citep{enoch11}, which is higher than the mass reported for the protostellar disk around, e.g., L1527 IRS \citep[0.007 $M_{\odot}$,][]{tobin12}, but of the same order as the disk in Lupus 3MMS \citep[0.1 $M_{\odot}$,][]{yen17}. 
If confirmed, the size and mass of the rotating structure in Ser-emb 1 are comparable to other previously detected protostellar disks around Class 0 protostars. 


The observed velocity gradient perpendicular to the outflow in, especially, C$^{18}$O, and the spin-up features apparent in its PV diagram 
are consistent with Keplerian rotation, similar to what has been observed for other Class 0 protostellar disks \citep[e.g., HH 212 in][]{lee17}. 
While 
a contribution from infalling envelope material cannot be excluded in Ser-emb 1, 
our PV diagrams are more comparable to  
the three-dimensional model of Keplerian rotation shown in \citet{oya16,oya17} for the COM emission in IRAS 16293A and L483, respectively, than to their corresponding models for infalling-rotating envelopes. 
In summary, this model predicts a spin-up feature \textit{only} in the PV diagram along the direction perpendicular to the outflow, and  symmetric, diamond-shaped emission for the PV diagram in the direction of the outflow for a Keplerian disk (similarly to what is observed for C$^{18}$O and H$_2$CO in Fig. \ref{pv}); while a model of an infalling-rotating envelope should present a spin-up feature in \textit{both} the PV diagrams prependicular to the outflow and along the direction of the outflow.  
However, better spatial resolution observations have recently ruled out the presence of a protostellar disk in L483, even when their infalling-rotating envelope model alone could not reproduce the features of the observed PV diagrams.
Higher spatial resolution and higher signal-to-noise ratio data are therefore needed to confirm the nature of the rotating structure in Ser-emb 1, and elucidate whether it is a pseudodisk or a fully rotationally supported disk).

The spatial coincidence of the compact hot corino emission and the candidate protostellar disk in Ser-emb 1 implies that the hot corino may be a warm disk. 
Including Ser-emb 1, four out of the ten detected hot corino sources are candidates to harbor rotationally supported protostellar disks (the others being IRAS 4A2, HH212, and IRAS 16293A, as explained in Sect. \ref{sec:intro}). 
Therefore, the hot corino chemistry in low-mass protostars could be located in the incipient protostellar disks formed as the result of the angular momentum conservation of the infalling envelope, as seen for the above mentioned sources in  \citet{oya16,lee17,codella18,lee19}, which would have important implications regarding the incorporation of precursors for prebiotic molecules to the planet-forming disks.

\section{Conclusions\label{conclusions}}

\begin{figure}[ht!]
\centering
\plotone{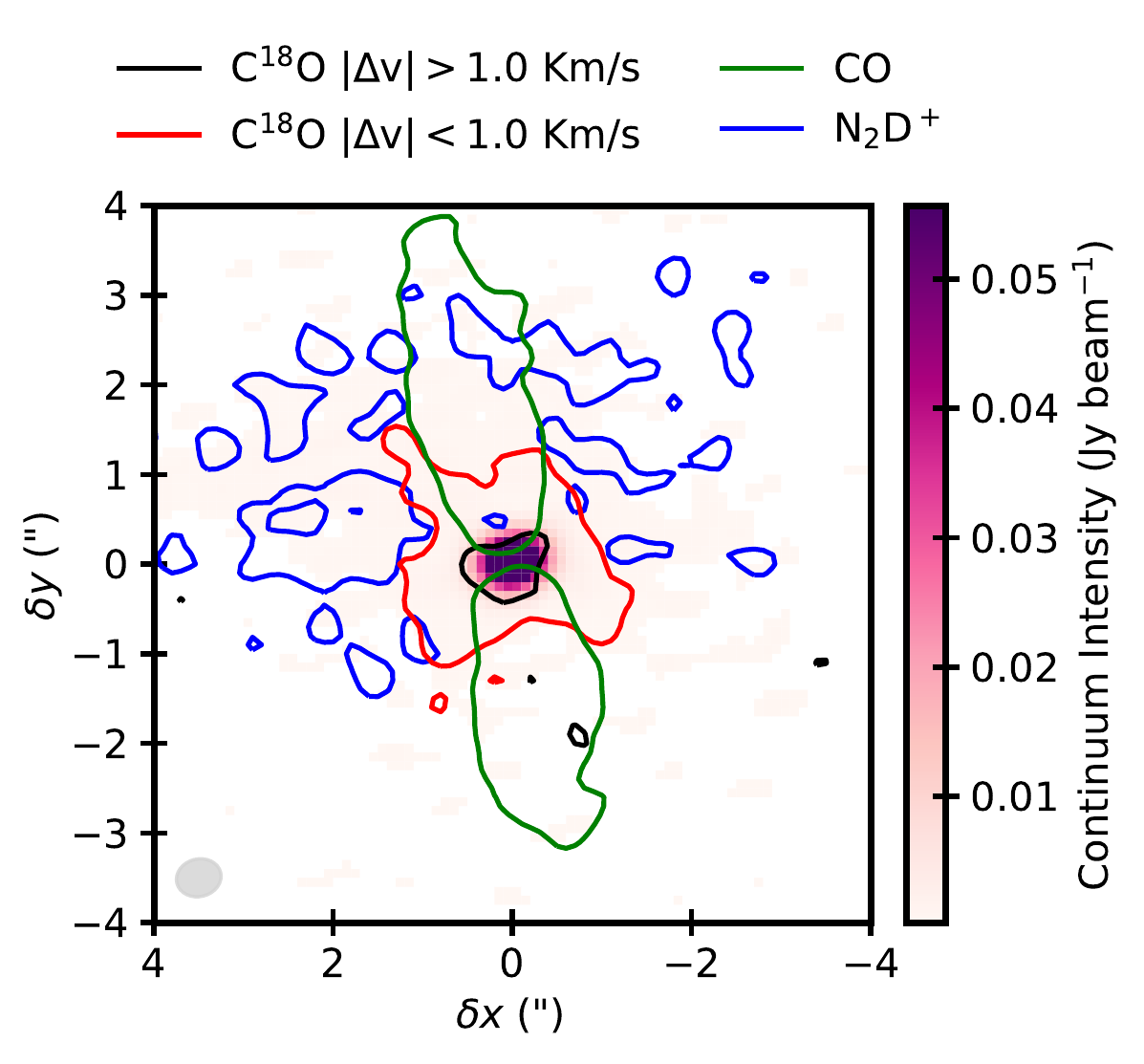}
\caption{Overview of the Ser-emb 1 source. The continuum map at 232 GHz traces the dust emission around the central protostar. The 6$\sigma$ contour of the CO 2 $-$ 1 moment 0 map is presented in green, tracing the bipolar jet and outflowt. The  5$\sigma$ contour of the moment 0 map encompassing the mid- and low-velocity channels (up to $\sim$1.0 km/s away from the $V_{LSR}$) of the C$^{18}$O 2 $-$ 1 line is shown in red, and trace the slow outflow; while the 4$\sigma$ contour high-velocity channels (from $\sim$1.0 to $\sim$1.5 km/s away from the $V_{LSR}$) in black trace the compact molecular emission. Finally, the 2$\sigma$ contours of the N$_2$D+ 3 $-$ 2 moment 0 map are shown in blue, and trace the cold outer envelope.  The size of the synthesized beam ($\sim$ 0.50$\arcsec$ $\times$ 0.42$\arcsec$) for the continuum map is shown in the lower left (see also Sect. \ref{app-obs})
\label{overviewcartoon}}
\end{figure}

\begin{enumerate}

\item The different species detected in our observational dataset trace chemically distinct structural components of the Ser-emb 1 YSO. These components can be observed in Fig. \ref{overviewcartoon}, and consist of
the central protostar, traced by the dust continuum map at 232 GHz; 
a rotating compact emission in the close vicinity of the central object, traced by the  C$^{18}$O 2 $-$ 1 velocity wings (black); a slow outflow traced by C$^{18}$O (red); 
a bipolar jet and outflow, traced by the CO 2 $-$ 1 emission (green);   
and the  cold outer envelope, traced by the N$_2$D$^{+}$ 3 $-$ 2 line (blue). 

\item In particular, we have detected ten chemically distinct species (and seven isotoplogs) with emission toward the continuum peak of Ser-emb 1 (Table \ref{lines1}), including the COM precursors H$_2$CO and CH$_3$OH, and two COMs: CH$_3$OCH$_3$, and CH$_3$OCHO. 

\item The rotational diagram analysis of the four CH$_3$OH lines results in a rotational temperature of  $T_{rot}$ = 261 $\pm$ 46 K.
The compact (radius $<$100 au) emission coming from CH$_3$OH, CH$_3$OCH$_3$, and CH$_3$OCHO, with a temperature above the ice sublimation temperature, implies that Ser-emb 1 harbors a hot corino. Additional unbiased observations are required to determine its full chemical richness. 



\item The high-velocity moment 0 maps of the C$^{18}$O 2 $-$ 1, and H$_2$CO 3$_{0,3}$ $-$ 2$_{0,2}$
transitions present spatially resolved blue- and red-shifted emission orthogonally to the outflow direction, indicative of rotation around the central protostar. 
The PV diagrams of the C$^{18}$O and H$_2$CO transitions along this direction and the direction of the outflow are consistent with a disk-like structure with radius $<$200 au, 
but contribution from the infalling-rotating envelope cannot be completely excluded. 

\item The spatial coincidence of the hot corino emission and the potential protostellar disk suggest that the hot corino could be actually a warm disk. 
Further observations with better spatial resolution and higher sensitivity of CH$_3$OH and other COMs are needed to evaluate this scenario.

\end{enumerate}

\acknowledgments

This work was supported by an award from the Simons Foundation (SCOL \# 321183, KO). J.B.B. acknowledges funding from the National Science Foundation Graduate Research Fellowship under Grant DGE1144152. The group of JKJ is supported by the European Research Council (ERC) under the European Union's Horizon 2020 research and innovation programme through ERC Consolidator Grant ''S4F'' (grant agreement No~646908). Research at Centre for Star and Planet Formation is funded by the Danish National Research Foundation.

%

\facility{ALMA}


\software{CASA (v4.3.0, v4.70, \& v5.3.0; \citep{mcmullin07}, MADCUBAIJ \citep{rivilla16a,rivilla16b}}



\appendix

\section{Observed spectral windows}\label{app-obs}

\begin{deluxetable}{cccc} 
\tablecaption{Spectral windows observed with the two correlator setups described in Sect. \ref{sec:observations}. For each spectral window, the first row corresponds to the channel rms and beam size reached after cleaning the data cube with a robustness parameter of 0.5; while the second row corresponds to the individually imaged transitions detected in the spectral windows, cleaned with \textit{robust} = 0.0. \label{spw}}
\tablehead{
\colhead{Frequency range} & \colhead{$\Delta$V} & \colhead{Channel rms} & \colhead{Beam size} \\
\colhead{(GHz)} & \colhead{(km s$^{-1}$)} & \colhead{(mJy beam$^{-1}$)} & \colhead{$^{\prime\prime}$}}
\startdata
217.209 $-$ 217.268 & 0.168 & 5.9 & 0.62 $\times$ 0.52\\ 
& & 7.5 & 0.54 $\times$ 0.48\\
218.193 $-$ 218.251 & 0.168 & 5.1 & 0.61 $\times$ 0.52 \\
& & 6.7 & 0.53 $\times$ 0.48\\
218.446 $-$ 218.505 & 0.168 & 5.0 & 0.61 $\times$ 0.52 \\
& & 6.4 & 0.53 $\times$ 0.48 \\
218.703 $-$ 218.762 & 0.167 & 5.1 & 0.61 $\times$ 0.52 \\ 
& & \nodata$^{(a)}$ & \nodata$^{(a)}$ \\
219.502 $-$ 219.619 & 0.167 & 6.3 & 0.62 $\times$ 0.50 \\
& & 7.0 & 0.53 $\times$ 0.46\\
220.340 $-$ 220.457 & 0.167 & 9.1 & 0.62 $\times$ 0.50 \\
& & 9.2 & 0.53 $\times$ 0.46 \\ 
230.479 $-$ 230.597 & 0.159 & 9.1 & 0.65 $\times$ 0.48 \\
& & 12.2 & 0.54 $\times$ 0.45\\
231.262 $-$ 231.380 & 0.158 & 6.8 & 0.58 $\times$ 0.47 \\
& &  \nodata$^{(a)}$ & \nodata$^{(a)}$ \\ 
231.481 $-$ 233.356 & 0.630 & 2.0 & 0.58 $\times$ 0.46 \\
& & 2.7 & 0.50 $\times$ 0.41 \\ 
242.978 $-$ 244.853 & 0.600 & 2.3 & 0.55 $\times$ 0.45 \\
& & 2.8 & 0.47 $\times$ 0.40 \\ 
244.164 $-$ 244.281 & 0.150 & 6.1 & 0.56 $\times$ 0.47 \\
& & 8.2 & 0.50 $\times$ 0.40 \\
244.877 $-$ 244.994 & 0.149 & 5.8 & 0.56 $\times$ 0.47 \\
& & 7.3 & 0.50 $\times$ 0.40 \\
258.098 $-$ 258.216 & 0.142 & 6.3 & 0.52 $\times$ 0.44 \\
& & 8.3 & 0.45 $\times$ 0.39\\
258.953 $-$ 259.070 & 0.141 & 5.8 & 0.51 $\times$ 0.44 \\
& & 7.5 & 0.45 $\times$ 0.39\\
260.459 $-$ 260.577 & 0.140 & 6.7 & 0.51 $\times$ 0.43 \\
& & 8.5 & 0.45 $\times$ 0.38\\
262.150 $-$ 262.267 & 0.140 & 7.3 & 0.51 $\times$ 0.43 \\
& & 8.4 & 0.44 $\times$ 0.38\\
\enddata
\begin{list}{}
\item $^{(a)}$No individual transitions were imaged with a robust = 0.0 in these spectral windows. 
\end{list}
\end{deluxetable}

\section{Rotational diagram}\label{app-rot}

Assuming local thermodynamical equilibrium (LTE) and optically thin lines, the integrated flux density $\int{S_{\nu} dv}$ of a given transition is related to the population of the corresponding upper level $N_u$ by Eq. \ref{nueq}:

\begin{equation}
N_u = \frac{4\pi\int{S_{\nu} dv}}{A_{ul}\Omega h c},
\label{nueq}
\end{equation}

where $A_{ul}$ is the Einstein coefficient of the transition (indicated in Table \ref{lines1} for every detected transition in our dataset), $\Omega$ is the solid angle subtended by the emitting region, $h$ is the Planck constant, and $c$ the speed of light. 

For the spectra in Figures \ref{ch3ohlines} and \ref{otherlines1}, $\Omega$ is assumed to be the beam solid angle.  
If the emitting region is smaller than the beam size, the observed integrated flux density is diluted in the beam. 
To correct for this beam dilution we have to apply the filling factor $\frac{\Omega_{beam}}{\Omega_{source}}$ to Eq. \ref{nueq}

The upper level population $N_u$ is related to the total column density $N_T$ of the species according to the Boltzmann equation:

\begin{equation}
\frac{N_u}{g_{up}} = \frac{N_T}{Q(T_{rot})} e^{-E_{up}/T_{rot}},
\label{rot}
\end{equation}

where $g_{up}$ and $E_{up}$ are the upper level degeneracy and upper level energy in K of the transition, respectively (also indicated in Table \ref{lines1} for the detected transitions), $T_{rot}$ the rotational temperature, and $Q(T_{rot})$ the partition function of the species (extracted from the JPL catalog). 
This way, every observed line represents one point in the rotational diagram of a species.  
Taking the natural logarithm of Eq. \ref{rot} leads to:

\begin{equation}
ln \frac{N_u}{g_{up}} = ln \frac{N_T}{Q(T_{rot})} - \frac{E_{up}}{T_{rot}}. 
\label{rotlin}
\end{equation}

Therefore, if $\frac{N_u}{g_{up}}$ of the different detected transitions of a given species are semi-log plotted against their upper level energies $E_{up}$, the rotational diagram can be fitted with a linear least squares regression, where $T_{rot}$ is the inverse of the slope and $N_T$ can be calculated from the intercept of the fit.

\end{document}